\documentclass[a4paper,utf8]{aa}

\usepackage{graphicx}
\usepackage{txfonts}
\usepackage[colorlinks, breaklinks, hyperindex, pdfpagelabels, hyperfootnotes=false]{hyperref}
\usepackage[decimalsymbol=period,mode=text,separate-uncertainty = true,multi-part-units = single]{siunitx}
\usepackage{caption}
\usepackage[labelfont={sf,bf,small},textfont={sf,small},justification=RaggedRight]{subcaption} 

\newcommand{\imgpath}{./figs/}
\newcommand{\plotpath}{./figs/}

\newcommand{\lcdatplotpath}{\plotpath}

\newcommand{\lcorbplotpath}{\plotpath}
\newcommand{\scorbplotpath}{\plotpath}
\newcommand{\lclcplotpath}{\plotpath}
\newcommand{\sclcplotpath}{\plotpath}
\newcommand{\scorbtwodplotpath}{\plotpath}

\newcommand{\corrdepthdepth}{0.026} 
\newcommand{\corrdepthstarptp}{0.0011} 
\newcommand{\corrdepthegress}{0.0019} 
\newcommand{\corrdepthsecond}{-0.012} 

\newcommand{\tsup}[1]{\textsuperscript{#1}}
\newcommand{\tsub}[1]{$_\mathrm{#1}$}

\newcommand{\eqnref}[1]{Eq.~(\ref{#1})\xspace}

\newcommand{\figref}[1]{Fig.~\ref{#1}\xspace}
\newcommand{\Figref}[1]{Figure~\ref{#1}\xspace}
\newcommand{\tabref}[1]{Table~\ref{#1}\xspace}

\newcommand{\secref}[1]{Sect.~\ref{#1}\xspace}
\newcommand{\Secref}[1]{Section~\ref{#1}\xspace}

\newcommand{\referee}{}

\newcommand{\includemplgraphics}[1]{\includegraphics[trim = 0 10 0 10, clip=true]{#1}}

\newcommand\tgtshort{KIC\,1255b\xspace}
\newcommand\tgtlong{KIC\,12557548\,b\xspace}
\newcommand\tgtstar{KIC\,12557548\xspace}
\newcommand\tgtstarshort{KIC\,1255\xspace}
\newcommand\keplersat{\emph{Kepler}\xspace}
\newcommand\sapqual{\texttt{SAP\_QUALITY}\xspace}
\newcommand\sapflux{\texttt{SAP\_FLUX}\xspace}
\newcommand\pdcsapflux{\texttt{PDCSAP\_FLUX}\xspace}

\newcommand\scshort{\textsc{SC}\xspace}
\newcommand\sclong{short cadence\xspace}

\newcommand\lcshort{\textsc{LC}\xspace}
\newcommand\lclong{long cadence\xspace}
\newcommand\Lclong{Long cadence\xspace}

\DeclareSIUnit\MJup{$M$\tsub{J}}

\newcommand\papertitle{Analysis and interpretation of 15 quarters of Kepler data of the \referee{disintegrating} planet \tgtlong}
\newcommand\papertitlerunning{Analysis and interpretation of the \referee{disintegrating} planet \tgtlong}

\hypersetup{
pdfauthor = {T.I.M. van Werkhoven et al.},
pdftitle = {\papertitle},
pdfsubject = {\papertitle},
pdfkeywords = {exoplanet, photometry, Kepler, eclipse},
pdfcreator = {pdfLaTeX with hyperref package},
pdfproducer = {pdfLaTeX},
colorlinks = true,
linkcolor = blue,         
citecolor = blue,         
filecolor = magenta,      
urlcolor = blue,          
}

\begin{document}

   \title{\papertitle}
   \titlerunning{\papertitlerunning}

   \subtitle{}

   \author{T.\,I.\,M.\ van Werkhoven \and M.\ Brogi \and I.\,A.\,G.\ Snellen \and C.\,U.\ Keller
          }

   \institute{Leiden Observatory, Leiden University,
              PO Box 9513, 2300 RA Leiden, The Netherlands\\
              \email{\url{werkhoven@strw.leidenuniv.nl}}
             }

	\date{Received: July 30, 2013 / Accepted: November 7, 2013}

 
  \abstract
   {\referee{The \keplersat object \tgtstar shows irregular eclipsing
     behaviour with a constant \SI{15.685}{\hour} period, but strongly
     varying transit depth. The object responsible for this is believed to be a disintegrating planet forming a trailing dust cloud transiting the star. A 1-D model of an exponentially decaying dust tail was found to reproduce the average eclipse in intricate detail. Based on radiative hydrodynamic modelling, the upper limit for the planet mass was found to be twice the mass of the Moon.}}
   {In this paper we fit individual eclipses, in addition to fitting binned light curves,
   to learn more about the process underlying the eclipse depth variation.
   Additionally, we put forward observational constraints that any model of this planet-star system will have to match.}
   {We manually de-correlated and de-trended \num{15} quarters of
     \keplersat data, three of which were observed in \sclong mode.
We determined the \referee{transit} depth, egress depth, and stellar intensity for each orbit and search for dependencies between these parameters.
We investigated the full orbit by comparing the flux distribution of a moving phase window of interest versus the out-of-eclipse flux distribution.
We fit \sclong data on a per-orbit basis using a two-parameter tail
model, allowing us to investigate potential dust tail property variations.}
   {We find two quiescent spells of $\sim$\num{30} orbital periods each where the transit depth is \referee{$<$}\SI{0.1}{\percent}, followed by relatively deep transits.
   Additionally, we find periods of on-off behaviour where $>$\SI{0.5}{\percent} deep transits are followed by apparently no transit at all.
   Apart from these isolated events we find neither significant
   correlation between consecutive transit depths nor a correlation between transit depth and stellar intensity.
   We find a three-sigma upper limit for the secondary eclipse of
   \num{4.9e-5}, consistent with a planet candidate with a
    radius of less than \SI{4600}{\kilo\meter}.
   Using the \sclong data we find that a 1-D exponential dust tail
   model is insufficient to explain the data.
   We improved our model to a 2-D, two-component dust model with an
   opaque core and an exponential tail.
   \referee{Using this model we fit individual eclipses observed in \sclong mode.}
   We find an improved fit of the data, quantifying earlier suggestions by \citet{budaj2013} of the necessity of at least two components.
   We find that deep transits have most absorption in the tail, and
   not in a disk-shaped, opaque coma, but the transit depth and the total absorption show no correlation with the tail length.}
   {}
   \maketitle
%

\section{Introduction}
\label{sec:introduction}

   \citet{rappaport2012} discovered the peculiar target \tgtstar in the \keplersat database, which shows dips in the light curve with a period of about \SI{15.7}{\hour} \referee{(constant to within \num{e-5}),} but a depth varying from less than \SI{0.2}{\percent} to up to \SI{1.3}{\percent}.
The phase-folded light curve shows no signs of ellipsoidal light
variations, which limits the mass of planet candidate \tgtlong\footnote{Henceforth denoted as \tgtshort.} to \referee{$<$}\SI{3}{\MJup}.
    
   \citet{rappaport2012} exclude several other scenarios for this
   target, including a dual-planet system and a low-mass eclipsing
   stellar binary. The authors argue for a disintegrating planet as
   the most likely scenario in which the close proximity to the host
   star causes parts of the planet's surface to evaporate. The
   evaporated gas drags dust along, which subsequently eclipses parts
   of the star. Because of the stochastic nature of this process, transits have variable depth. The authors also find evidence of forward scattering due to this dust cloud, which creates a slight increase in intensity just before ingress. This scenario puts an upper limit on the planet's escape velocity, such that a Mercury-mass planet is the most likely candidate. The authors qualitatively investigate the likeliness of this scenario and find it to be consistent with observations.
  
  Subsequently, \citet{brogi2012} quantitatively investigated the
  planet hypothesis using a 1-D model to constrain the transit
  parameters, the shape of the dust cloud, and the average particle size. 
They find that a dust cloud with $\sim$\SI{0.1}{\micro\meter}-sized
particles \referee{best matches} the observed, average eclipse light curve. 
The authors also find different system parameters for subsets of the transits consisting of relatively shallow (\SI{0.2}{\percent} to \SI{0.5}{\percent}) and deep ($>$\SI{0.8}{\percent}) eclipses.

\citet{perezbecker2013} investigate the evaporation dynamics of this planet candidate and, through radiative hydrodynamic modelling, argue that it is losing mass at a rate of $\dot{M} \ga 0.1\,M_\oplus$\,Gyr\tsup{-1}.
They conclude that the planet candidate has a mass of less than $\num{0.02}\,M_\oplus$, or twice the mass of the Moon.
According to the authors, \tgtshort may have lost up to \SI{70}{\percent} of its initial mass, with only the inner iron core left. 
\referee{\citet{budaj2013} investigates the dust tail properties in more detail and} argues that the particle size changes along the tail, where micron-sized particles best explain ingress while \numrange{0.1}{0.01}-micron sized particles fit egress best.
  
Here we investigate the shape of individual eclipses and provide statistical constraints on the system.
We extend the previous model from \citet{brogi2012} to a pseudo \referee{2-D}
variant where the vertical extent of the cloud of dust is not
neglected, and an opaque core is included as a disk centred on the planet candidate.
The \sclong data for quarters \num{13} through \num{15} allow us to
fit the model on a per-transit basis to compare individual transits.
Using the \num{15}-quarter coverage of the target we investigate
correlations between the transit depth, the depth at egress, and the
stellar activity as well as variations of these parameters.
Additionally, we derive a three-sigma upper limit for the secondary
eclipse of \num{4.9e-5}, which is consistent with an object radius smaller than \SI{4600}{\kilo\meter} \referee{for an albedo of 1}.
  
The data reduction is explained in \secref{sec:datareduction}, and the light curve analysis is presented in \secref{sec:kic1255banalysis}.
\Secref{sec:model} describes the improved model with which we
investigate per-orbit properties for the \sclong data as well as \referee{an} analysis and interpretation of those results.

\section{Data reduction}
\label{sec:datareduction}

\keplersat is a \SI{0.95}{\meter}-aperture Schmidt telescope with a \SI{16}{\degree} diameter field of view, observing \num{156453} stars with a \num{95} megapixel, \num{42} science CCD focal plane \citep{koch2010}.
The telescope was launched on 6 March 2009 and started observing on 2
May 2009.
Unfortunately, in May 2013 \keplersat went into safe mode due to a second reaction wheel failing.

At the basic level, frames are recorded by integrating for \SI{6.02}{\second}, followed by a \SI{0.52}{\second} readout.
Data is then stored in \sclong (\scshort) mode by \referee{co-adding} \num{9} frames, giving a cadence of \SI{58.86}{\second} with a \SI{54.18}{\second} exposure time \citep{gilliland2010} or in \lclong (\lcshort) mode by \referee{co-adding} \num{270} frames resulting in a cadence of \SI{29.4244}{\minute} and an exposure time of \SI{27.1}{\minute} \citep{jenkins2010b}.
Because of the limited \referee{telemetry bandwidth}, \keplersat observes \referee{no more than} \num{512} targets -- about \SI{0.3}{\percent} of the total -- in \scshort mode.

At the time of writing, \num{15} quarters of \keplersat data are publicly available at the Multi-Mission Archive (MAST\footnote{\url{http://archive.stsci.edu/kepler/data_search/search.php}}) at the Space Telescope Science Institute. 
\tgtstar was observed in \lcshort mode for quarters \numrange{1}{12}, and in \scshort mode during the last quarters (\numrange{13}{15}). 
The latter data have a \num{30} times shorter exposure time, resulting in a \referee{factor of} $\sqrt{30} \approx \num{5.5}$ lower signal-to-noise ratio, assuming pure Poisson noise.

The \keplersat pipeline \citep{jenkins2010} delivers the \referee{light curve} as \emph{Simple Aperture Photometry} flux (\sapflux), as well as an automatically reduced \emph{Presearch Data Conditioning} \referee{(PDC)} flux (\pdcsapflux). 
The PDC aims to remove systematic errors from the raw flux time series.
Since \tgtstarshort exhibits a very peculiar light curve, the pipeline
may have difficulty in automatically reducing the data. 
Indeed, we find that for the \lcshort data, deep transit data are flagged as outliers.
The automatically reduced \scshort data has a noise level approximately \num{1.5} times \referee{higher} than our manually reduced data.
Because of this, we started the data reduction from the raw \sapflux and manually de-correlate the data.

The data reduction is explained in more detail below. First we selected usable data and filtered out bad data. For \sapflux only, we manually de-correlated the signal using linear co-trending basis vectors supplied by the \keplersat pipeline to remove systematics from the signal. Finally, we removed the stellar signal by de-trending the flux.

\subsection{Data selection}
\label{subsec:dataselection}

We used all \lcshort data from \keplersat quarters \num{1} through \num{15}, and all \scshort data, which were taken during quarters \numrange{13}{15}.
We ignored non-finite data and most non-zero \sapqual flagged data. 
The \keplersat pipeline erroneously marks some data with flag \num{2048} (``Impulsive outlier removed after cotrending'', \citep[p.~20]{fraquelli2012} for transits deeper than $\sim$\SI{0.8}{\percent}.
Considering \referee{that} this \referee{is a} highly variable target, it is not unlikely that the automated \keplersat pipeline has some difficulties, and indeed the \keplersat archive manual warns of these cases (\texttt{kepcotrend} documentation\footnote{\url{http://keplergo.arc.nasa.gov/ContributedSoftwareKepcotrend.shtml}.}).
Ignoring these data points would not correctly sample deep transits, and we therefore included these flagged data.
We have not found any other \sapqual flags that correlate with the
orbital phase, and we therefore removed all other flagged data from our
analysis.

\begin{figure*}[!htb]
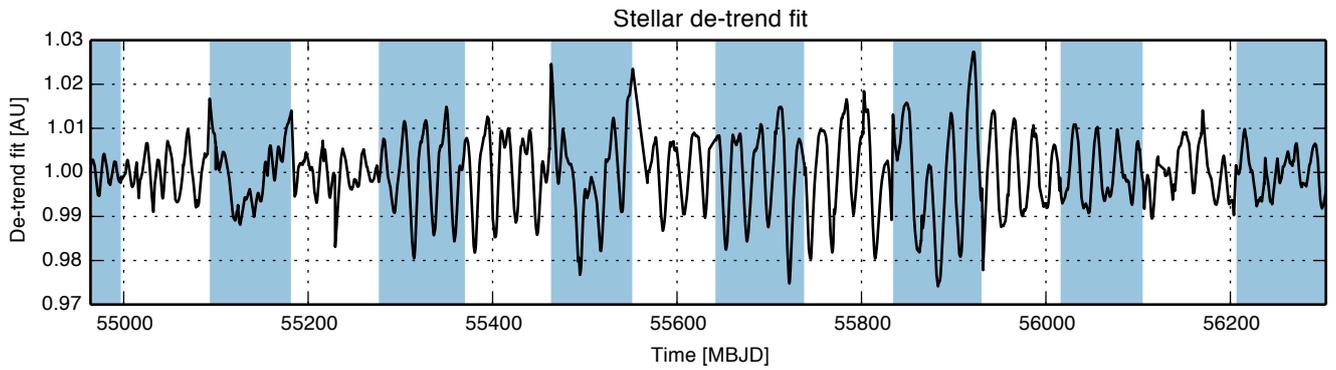

  \centering
  \includemplgraphics{\lcdatplotpath KIC1255b_15Q_lc_I_re2plot_SAP_detrend.pdf}
  \caption{\referee{De-trending fit for the de-correlated, \lclong \sapflux}, showing the stellar variability during quarters \num{1} through \num{15}.
  Odd quarters are shaded.
  \label{fig:detrend_fit}}
\end{figure*}

\subsection{De-correlating systematics}
\label{subsec:decorr}

To remove systematics from the \referee{light curve}, the \keplersat archive provides a set of \num{16} linear co-trending basis vectors (LCBV) for each detector, which are derived from a subset of highly correlated and quiet stars and are meant to remove satellite systematics from the data \citep[p.~21]{fraquelli2012}.
The pipeline automatically de-correlates the \referee{light curve} for all targets against these LCBVs, but the \keplersat archive manual recommends to manually de-correlate signals that are highly variable. 

Linear co-trending basis vectors are only available for \lcshort data and cannot be constructed manually for the \scshort data because not enough targets are observed in \scshort to generate a set of LCBVs.
To de-correlate the \scshort data, we linearly interpolated the LCBVs on the \scshort time points before de-correlation.
We successfully used this technique to minimise the out-of-eclipse residual variance of the \scshort data of quarters \numrange{13}{15}.
We note that this approach is unable to correct for systematics occurring on timescales significantly shorter than \SI{29}{\minute} (i.e.\ that of the \lclong). 

\begin{figure}
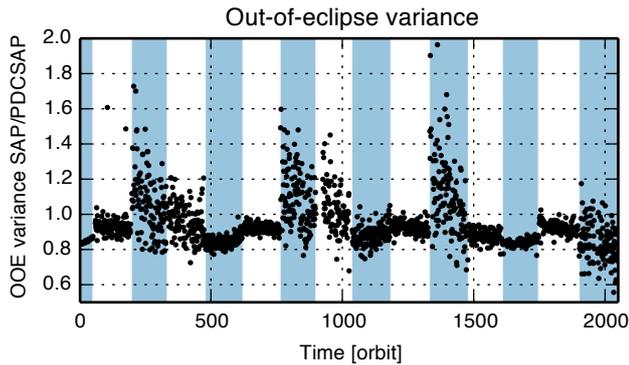

  \centering
  \includemplgraphics{\lcdatplotpath KIC1255b_15Q_lc_I_re2plot_ooe_variance_ratio.pdf}
  \caption{Ratio of \lclong \sapflux over \pdcsapflux standard
    deviation during the out-of-eclipse part of the orbit (phase $\varphi \in [-0.4, -0.1]; [0.2, 0.4]$).
    Odd quarters are shaded.
    The flux variance differs per quarter.
  \label{fig:ooe_variance}}
\end{figure}

We \referee{de-correlated} the flux as follows.
First we interpolated the \keplersat LCBVs on \sapflux exposure times
for the \scshort data.
Then we excluded the primary eclipse at orbital phase $\varphi \in [-0.15, 0.20]$ from the fitting process.
We least-squares fit all \num{16} vectors to the remaining data and selected the first $n$ vectors that reduce the out-of-eclipse variance significantly (see \figref{fig:ooe_variance}).
In our case, we used $n$ = \num{2} vectors for the \scshort data and $n$ = \numrange{2}{5} for the each of the \lcshort data quarters.

\subsection{Removing stellar variability}
\label{subsec:detrend}

When the de-correlation is performed correctly, it must preserve all astrophysical signal.
This includes stellar variability, which has to be removed to analyse the eclipse behaviour.
Since the data has jumps in both time and flux due to gaps in the
data, we de-trended the data piece-wise per block, where a block is
delimited by either a jump in time or flux.
A time jump is defined as a gap in data of more than twice the data
cadence, a flux jump is defined as a change of flux of more than
\SI{0.5}{\percent} (\referee{\lcshort}) and \SI{2}{\percent} (\referee{\scshort}) difference between
two consecutive data points, and if the difference in the median of
the \num{20} data points around the jump differs by more than \num{3.5} times the standard deviation of those points.
These parameters were chosen on trial and error basis.

For each block, we masked out the transits during orbital phase $\varphi \in [-0.15, 0.20]$ when fitting.
For blocks up to \num{10} data points or spanning a single orbit, we normalised the data by the median. 
For blocks up to \num{200} data points we normalised using a second-order polynomial fit, and for larger blocks we fit a cubic spline to the data with a knot at each orbit at $\varphi = \num{0.5}$\referee{, exactly opposite to the transit}.
We flagged data that are near the edge of a block: within one orbital period of the edge, or data outside the outer knots of a spline fit.
Both of these flags are excluded in subsequent analyses since these
data are poorly fitted and may introduce \referee{unwanted} errors in subsequent steps.
By removing the stellar variation with only one degree of freedom per orbit (i.e.\ a spline knot), we keep the transit signal intact.

\Figref{fig:detrend_fit} shows the \referee{de-trending} fit by which we divided the de-correlated \sapflux to remove the stellar signal.
The variability in this plot is caused by star spots coming in and out of view due to the stellar rotation.
We find a period for the stellar rotation of \SI{22.65(5)}{\day} by auto-correlating the signal, \referee{which is consistent with the findings of \citet{kawahara2013}.}

\subsection{Orbital parameters}
Once the signal is de-correlated and the stellar variability is
removed, we computed the orbital period of the planet candidate by minimising the difference between the phase-folded \lcshort data and the best-fitting model from \citet{brogi2012} in a least-squares sense.
This method is similar to phase dispersion minimisation \citep{stellingwerf1978} with the exception that we use a model instead of smoothed data.
For these analyses, we used barycentric Julian days expressed in barycentric dynamical time, as given by the \keplersat pipeline \citep{eastman2012}.
The values for the period with \referee{one-sigma} uncertainties are listed in \tabref{tab:orbitperiod}.
We do not find significantly different values for the period for \pdcsapflux and \sapflux reduced data.
\begin{table}
\centering
  \renewcommand{\arraystretch}{1.2}
  \sisetup{separate-uncertainty = false}
    \caption{\referee{Comparison between the literature values of the orbital period of \tgtshort and the value determined through our analysis.}}
    \begin{tabular}{ll}
    Source & Period (d)
     \\
    \hline
    \citet{rappaport2012} & \num{0.65356 (1)}
     \\
    \citet{budaj2013}     & \num{0.65355 21(15)}
    \\
    This analysis            & \num{0.65355 38(1)}
    \\
  \end{tabular}
  \label{tab:orbitperiod}
\end{table}
Additionally, we fitted the \referee{1-D} model parameters described in \citet{brogi2012} using \num{15} quarters of data (see \secref{subsec:model_lc} and \secref{subsec:fit_lc} for more details).
Likewise, we calculated these parameters for deep (more than \SI{0.8}{\percent}) and shallow transits (\SI{0.2}{\percent} to \SI{0.5}{\percent}) as well as all transits.
The results are shown in \tabref{tab:orbitparms}.
\def\tol#1#2#3{\hbox{\rule{0pt}{1.1em}$\num{#1}\,^{\num[retain-explicit-plus]{#2}}_{\num{#3}}$ }}
\def\toln#1#2#3{\makebox[0pt][r]{$-$}\hbox{\rule{0pt}{1.1em}$\num{#1}\,^{\num[retain-explicit-plus]{#2}}_{\num{#3}}$ }}
\begin{table}
\centering
  \renewcommand{\arraystretch}{1.2}
  \sisetup{separate-uncertainty = false}
    \caption{Best-fit \referee{1-D} model parameters and their one-sigma uncertainties, as derived from the Markov-chain Monte Carlo (MCMC) analysis.
    From top to bottom: impact parameter $b$, mid-transit phase offset $\Delta\varphi_\mathrm{0}$, decay factor $\lambda$, total extinction cross-section (in units of stellar area) $c_\mathrm{e}$, asymmetry parameter $g$, and single-scattering albedo $\varpi$.
    }
    \begin{tabular}{llll}
    Parameter & Average & Deep & Shallow \\
    \hline
    $b$           & \tol{0.63}{+0.022}{-0.023}		
                  & \tol{0.51}{+0.031}{-0.022}       
                  & \tol{0.62}{+0.030}{-0.029} \\    

    $\Delta\varphi_\mathrm{0}\times\num{e3}$
                      & \toln{1.35}{+0.43}{-0.44}		
                  & \toln{1.01}{+0.55}{-0.51}       
                  & \tol{0.27}{+0.66}{-0.80}\\       

    $\lambda$     & \tol{5.83}{+0.21}{-0.19}		
                  & \tol{5.84}{+0.23}{-0.24}         
                  & \tol{4.80}{+0.24}{-0.20}\\       
    $c_\mathrm{e}$& \tol{0.0227}{+0.0020}{-0.0013}	

                  & \tol{0.0415}{+0.0048}{-0.0035}   

                  & \tol{0.0139}{+0.0009}{-0.0007}\\ 

    $g$           & \tol{0.809}{+0.033}{-0.045}		
                  & \tol{0.810}{+0.026}{-0.039}      
                  & \num{0.809} (fixed) \\
    $\varpi$      & \tol{0.49}{+0.090}{-0.088}		
                  & \tol{0.96}{+0.12}{-0.11}         
                  & \num{0.49} (fixed) \\
  \end{tabular}
   \label{tab:orbitparms}
\end{table}

\section{\tgtshort analysis}
\label{sec:kic1255banalysis}

\begin{figure}[htb]
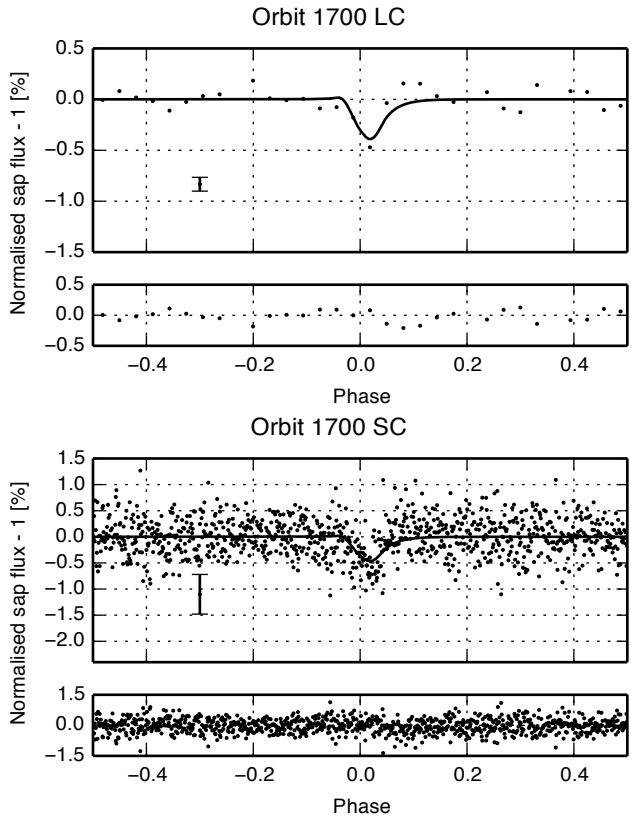

  \centering
  \begin{subfigure}[b]{\linewidth}
    \includegraphics{{{\lclcplotpath sap_lc_orbit_resid_1700_tmid56074.9}.pdf}}
  \end{subfigure}
  \begin{subfigure}[b]{\linewidth}
    \includegraphics{{{\sclcplotpath sap_sc_orbit_resid_1700_tmid56074.9}.pdf}}
  \end{subfigure}
  \caption{Orbit \num{1700} of \tgtshort in long- (\emph{top}) and \sclong (\emph{bottom}) data.
  The \sclong data has a $\num{30}\times$ higher temporal \referee{resolution}, but
  correspondingly higher noise.
  \referee{We overplot \referee{the} best-fitting 1-D model in the upper panels.}
  \label{fig:lc_one_orbit}}
\end{figure}

After reducing all available data, we investigated them on a per-orbit basis.
\referee{We numbered the orbits} sequentially; orbit \num{1} is the first orbit observed by \keplersat and \num{2050} is the last orbit \referee{in these data}.
We used light curves for \num{1773} orbits in total, excluding bad data as identified by the \keplersat \referee{pipeline} (i.e.\ \sapqual) as well as data that was poorly de-trended.
\referee{As an example} orbit \num{1700} is shown in \figref{fig:lc_one_orbit} for both the \lclong (\lcshort) and \sclong (\scshort) data.
\subsection{Primary eclipse}
\label{subsec:primeclipse}

Since the \lcshort data has a cadence of \SI{29.4}{\minute}, we have an average of \num{31.984} data points per \SI{15.6854}{\hour} orbit and are limited to about $\num{31.984}\times\num{0.1}\approx\num{3}$ data points per individual transit (see for example \figref{fig:lc_one_orbit}, top panel).
Hence we only fit the depth of the transits using the 1-D model with the best-fitting parameters of \citet{brogi2012}.
To measure the depth, we performed a least-squares fit using a scaled best-fitting model to the data at $\varphi \in [-0.1, 0.2]$.

For \lcshort data, we convolved the model data with the \keplersat exposure time \referee{before fitting}, as explained in \citet{brogi2012}.
For the \scshort data we did not convolve the data with the \keplersat
exposure time, and there is sufficient temporal resolution to
additionally fit the onset of the transit. The transit onset is
determined by shifting the best-fitting model in time.
A least-squares \referee{fit} yields both onset and depth simultaneously for each transit.

\begin{figure}[t]
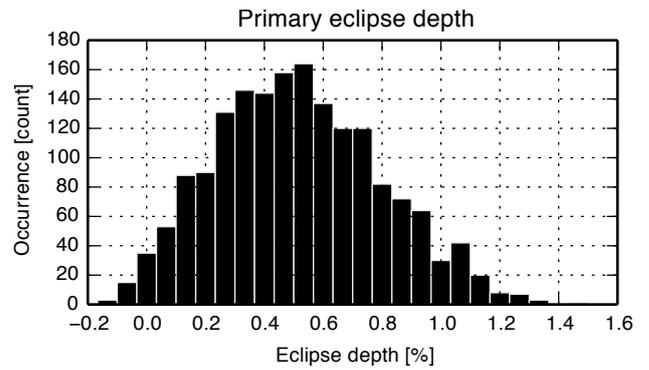

  \centering
  \includemplgraphics{\lcorbplotpath sap_Primary_eclipse_depth_histo.pdf}
  \caption{All \lclong transit depths obtained from scaling \referee{the} best-fitting model to \sapflux data.
  Here we plot the depth at the minimum of those scaled models.
  The few negative eclipse depths are due to noise in the data.
  \label{fig:lc_depthhisto}}
\end{figure}
A histogram of all \lcshort transit depths is shown in \figref{fig:lc_depthhisto}.
\Figref{fig:sapbinned} shows the normalised, binned, phase-folded
light curve for \sclong and \lclong data of transits of depth
\SIrange{0.8}{1.0}{\percent}, with the residuals 
between the model and the data shown below.
The model deviates from the data during egress, showing that the
simple, exponential tail model cannot explain the light curve in full detail.
\begin{figure}[!th]
  \centering
  \begin{subfigure}[b]{\linewidth}
    \includegraphics{{{\lcorbplotpath sap_lc_binned_normd_nbin192_d0.8_1.0_with_residual2_model}.pdf}}
  \end{subfigure}
  \begin{subfigure}[b]{\linewidth}
    \includegraphics{{{\scorbplotpath sap_sc_binned_normd_nbin192_d0.8_1.0_with_residual2_model}.pdf}}
  \end{subfigure}
  \caption{Phase-folded, normalised flux binned in \num{192} bins for \lclong (\emph{top}) and \sclong (\emph{bottom}) data for transits with depths from \SIrange{0.8}{1.0}{\percent}.
  The number of data points in each bin is indicated with horizontal
  bars according to the right y-axis.
  The model is based on the parameters from the `deep' column in \tabref{tab:orbitparms}.
  The best-fitting model deviates from the \sclong data during both ingress and egress.
  The vertical bar indicates the median three-sigma error of the binned flux.
  \referee{The residual RMS for these data in the phase displayed are \num{2.0e-4} and \num{3.7e-4} respectively, showing a greater mismatch between the \sclong data and the 1-D model.}
  \label{fig:sapbinned}}
\end{figure}

\begin{figure*}
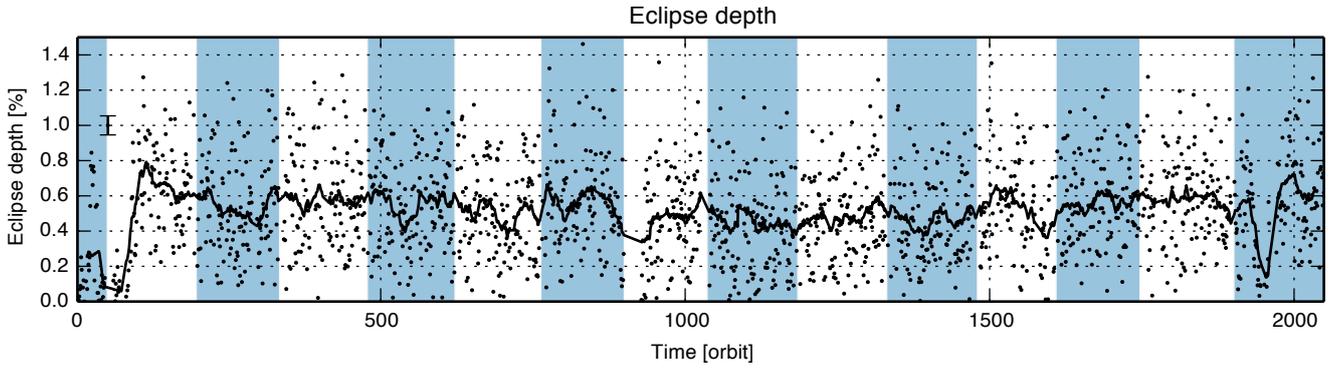

  \centering
  \includemplgraphics{\lcorbplotpath sap_pri_ecl_depth_evol_smooth_n30_wide.pdf}
  \caption{Transit depth as a function of orbit for \lclong data.
  The solid line is a \num{30}-orbit moving average.
  There are two quiet regions around orbit \num{50} and \num{1950}.
  Odd quarters are shaded.
  \label{fig:lc_depthevol}}
\end{figure*}

\begin{figure}
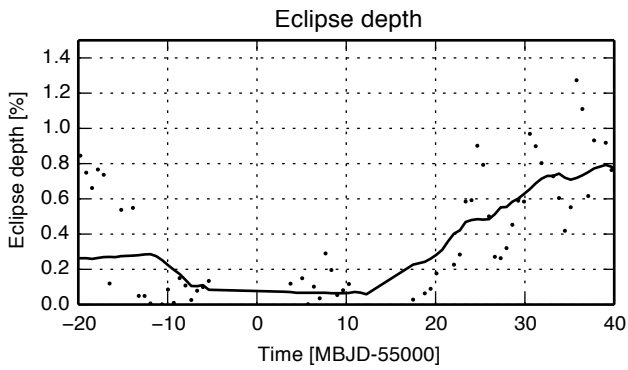

  \centering
  \includemplgraphics{\lcorbplotpath sap_pri_ecl_depth_evol2_smooth_quiet.pdf}
  \caption{\referee{Transit depth as a function of orbit, showing the first quiet period around orbit \num{50} in \figref{fig:lc_depthevol} in more detail.}
  \label{fig:lc_depthevol_quiet}}
\end{figure}

\begin{figure}
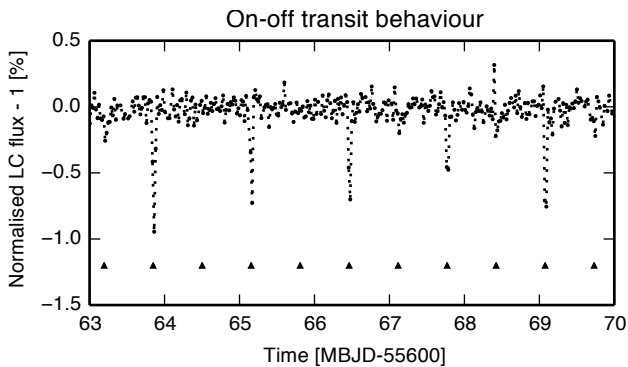

  \centering
  \includemplgraphics{\plotpath kic1255b_LC_on-off_period_orb1068--1084-time.pdf}
  \caption{\referee{\Lclong flux as a function of time where \tgtshort shows on-off-like behaviour in the transit depth.
  The triangles indicate the midpoint of the transits.}
  \label{fig:lc_depthevol_onoff}}
\end{figure}

In \figref{fig:lc_depthevol} we plot the transit depth as a function
of orbit, with a \num{30}-orbit (approximately one stellar rotation)
moving average plotted as a solid black line.
There are two quiet regions around orbit \num{50} and \num{1950} (MBJD \num{55000} and \num{56250}) during which the average transit depth is on the order of \SI{0.1}{\percent}\referee{, the former is plotted in \figref{fig:lc_depthevol_quiet}}.
The quiet periods are followed by periods during which the moving-average depths are \SI{0.8}{\percent} and \SI{0.7}{\percent}, approximately \SIrange{0.1}{0.2}{\percent} deeper than the average.
Additionally, we find times at which the transits appear in an on-off-pattern where $>$\SI{0.5}{\percent} deep transits are followed by apparently no transit \referee{signal} at all for up to \num{11} orbits.
This occurs for example around orbit \num{1076} and to a lesser extent around orbit \num{940} (MBJD \num{55667} and \num{55578}).
\referee{In \figref{fig:lc_depthevol_onoff} we show the period around orbit \num{1076}.}

\subsubsection{Transit depth correlation}

\citet{rappaport2012} argue that the dust has a sublimation lifetime of \SI{3e4}{\second}, or \SI{8}{\hour}, such that it does not survive one orbit.
To test this, we investigated the correlation between consecutive transit depths, as well as transit depth and consecutive egress depth, defined as the depth \referee{during} $\varphi \in [0.055, 0.15]$.
In the former case we expect a correlation if the dust generation
lasts longer than one orbit, such that \referee{subsequent} transits are correlated.
When investigating the transit and consecutive egress depths, we
quantified the longevity of a dust cloud, such that deep transit
clouds survive as an elongated tail in the next transit, which would
lead to a particularly long egress signature.
In this scenario, we would observe a deep transit as caused by a recent outburst where the dust is close to the planet candidate, eclipsing a large part of the star.
Under the influence of gravity and the stellar radiation pressure this
cloud would deform into a comet-like tail during the orbit, such that
a more tenuous dust tail would eclipse the star during the \referee{following} orbit.

\begin{figure}
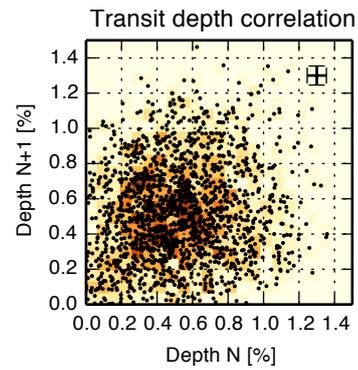

  \centering
  \includemplgraphics{\lcorbplotpath sap_pri_ecl_depth_d1_corr.pdf}
  \caption{Transit depth plotted against the following transit depth, showing no correlation ($R^2=\num{\corrdepthdepth}$).
  This constrains the dynamic processes that underlie the dust cloud generation.
  The lower-left corner is slightly overpopulated due to two quiet periods of \tgtshort as explained in \secref{subsec:primeclipse}.
  The cross indicates the median error for the depth estimate.\label{fig:seq_depth_corr}}
\end{figure}

For this analysis we selected all pairs of sequential orbits.
We plot the depth \referee{versus} the depth and egress depth \referee{for} consecutive orbits in the \lcshort data in \figref{fig:seq_depth_corr} and \ref{fig:lc_depth_vs_egress1}\referee{, respectively}
\begin{figure}[thb]
  \centering
  \includemplgraphics{\lcorbplotpath sap_pri_ecl_egress_depth_d1_corr.pdf}
  \caption{Transit depth versus egress intensity for consecutive orbits
    for the \lclong data. We observe no significant correlation ($R^2=\num{\corrdepthegress}$).
    \label{fig:lc_depth_vs_egress1}}
\end{figure}

There is no obvious correlation between the depth of consecutive transits (Pearson's $R^2 = \num{\corrdepthdepth}$) nor between depth and consecutive egress depth ($R^2 = \num{\corrdepthegress}$).
The absence of a correlation between consecutive transit depths
indicates that the process underlying the dust generation is erratic and occurs on time scales shorter than one orbit.
Furthermore, \referee{the} lack of correlation between transit and consecutive egress
depths is consistent with earlier findings by \citet{rappaport2012}
and puts an upper limit on the dynamical time scale of the dust tail dissipation at one orbit, i.e.\ \SI{15.7}{\hour}.
\referee{This is consistent with \citet{perezbecker2013}, who calculated the dynamical timescale of the dust tail and found it to be approximately \SI{14}{\hour}.}

We also investigated the correlation between the stellar intensity in
the \keplersat band and the transit depth. The intensity is a proxy
for the stellar activity, and a correlation might reveal the influence of stellar activity on the transit depth and thus gas and dust generation, as observed for Mercury \citep{potter1990}.
As we observe the rotational modulation of the stellar intensity
due to star spots, and not the absolute intensity, the amplitude of
the cyclical intensity variations is also influenced by the spatial
and size distributions of the star spots. As is observed on the Sun,
we assume that there is a positive correlation between the amplitude
of the cyclical intensity variations and the magnetospheric effects
affecting the planet and its dust tail.

\begin{figure}
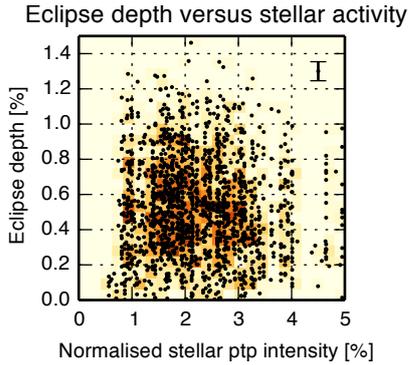

  \centering
  \includemplgraphics{\lcorbplotpath sap_pri_ecl_depth_vs_stellar_ptp_hist.pdf}
  \caption{Transit depth versus stellar activity, as determined by the peak-to-peak value of the stellar intensity in a \SI{24}{\day} moving window centred at the time of the transit. 
  \label{fig:depth_stellar_corr}}
\end{figure}

\referee{We measured the stellar variability as a by-product of our data de-trending}, as described in \secref{subsec:detrend}.
From the stellar variability we computed the peak-to-peak value in a moving \SI{24}{\day} window (\referee{about} one stellar rotation period) to obtain a proxy for the stellar activity, where we assume that a higher amplitude corresponds to a more active star.
We show the transit depth versus the stellar activity at each orbit in \figref{fig:depth_stellar_corr}.
We observe no significant correlation between transit depth and
our stellar activity proxy ($R^2$ = \num{\corrdepthstarptp}).
\referee{\citet{kawahara2013} performed a time-series analysis of the transit depth evolution and found a periodicity close to the rotation period of the star, which they interpret as an influence of stellar activity on the atmospheric escape of the planet candidate.
We have not found any evidence for this using our method.}

\subsection{Secondary eclipse}
\label{subsec:sececlipse}

\begin{figure}[!hbt]
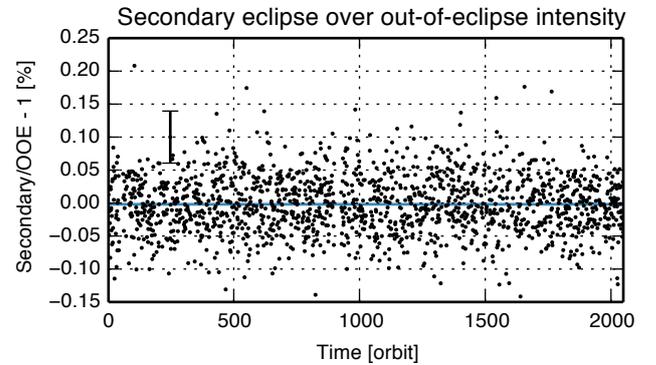

  \centering
  \includemplgraphics{\lcorbplotpath sap_sec_ecl_vs_ooe_ratio.pdf}
  \caption{\Lclong flux during \referee{the} secondary eclipse divided by flux
    during out-of-eclipse \referee{periods} as a function of orbit.
  The weighted-average three-sigma upper limit is \num{4.9e-5}.
  The black error bar is the mean error of the individual data points.\label{fig:lc_sececlipse_ratio}}
\end{figure}

We divided the \lcshort flux during the expected secondary eclipse ($\varphi \in [0.45, 0.55]$) by the out-of-eclipse flux.
Using this ratio ensures that any residual deviations from unity in
the flux due to inaccuracies in the de-trending will not be mistaken
for a real signal at the time of the secondary eclipse.
This ratio is plotted in \figref{fig:lc_sececlipse_ratio}, with the mean error of the data points shown on the left.
The weighted average and \referee{the} error of all data give a three-sigma upper limit for the secondary eclipse of \num{4.9e-5}.
Using simple geometry we find that a planet candidate with radius
smaller than \SI{4600}{\kilo\meter}, or an Earth-sized object with an albedo of $\sim$\num{0.5} is consistent with this finding.
Furthermore, we find no correlation ($R^2$ = \num{\corrdepthsecond})
between the depth of the primary eclipse and the secondary eclipse, indicating that even deep primary eclipses do not leave any significant, back-\referee{scattering} dust after half an orbit.

\subsection{Full orbit}
\label{subsubsec:fullorbit}

\begin{figure}[thb]
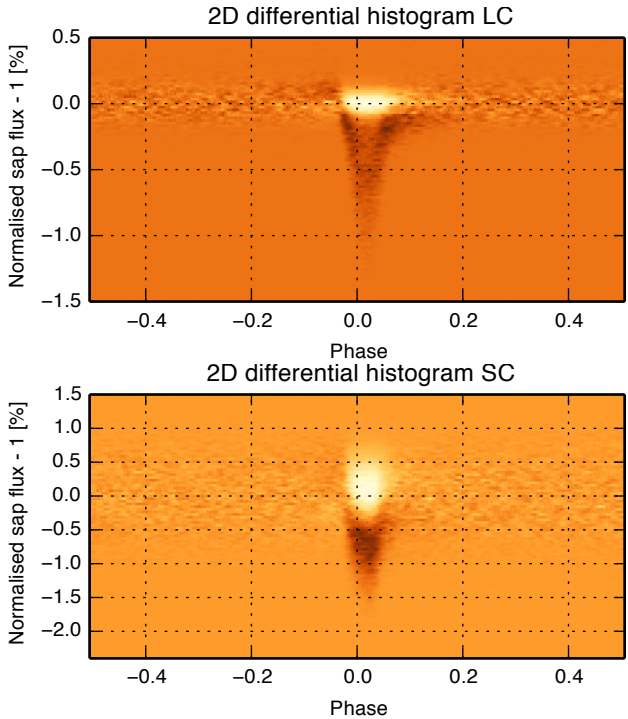

  \centering
  \begin{subfigure}[b]{\linewidth}
    \includemplgraphics{\lcorbplotpath sap_lc_2d_diff_histo_d-1-5_1x4_100ibin_80phbin_diff.pdf}
  \end{subfigure}
  \begin{subfigure}[b]{\linewidth}
    \includemplgraphics{\scorbplotpath sap_sc_2d_diff_histo_d-1-5_1x4_100ibin_80phbin_diff.pdf}
  \end{subfigure}
  \caption{2-D differential histogram comparing the flux distribution at a certain phase against the out-of-eclipse flux distribution for all long- (\emph{top}) and \sclong (\emph{bottom}) data.
  Excess flux compared to the out-of-eclipse histogram shows up as
  bright areas.
  We use \num{100} intensity bins and \num{320} phase bins of width \num{0.0125} such that the phase bins are $\num{4}\times$ oversampled.
  Note the excess flux around phase $\varphi$ = \num{-0.1} in the
  \lclong data indicating the forward scattering, which is not visible in the \scshort data.
  \label{fig:2ddiffhisto}}
\end{figure}

To investigate features at phases other than the transit, we compared the flux distribution at different phases against the distribution of the out-of-eclipse flux.
Since we expect a flat light curve during the out-of-eclipse phase, significant deviations from zero in the differences between these distributions are indications of potential features due to the dust.
We plot the difference between the out-of-eclipse distribution and a distribution of the flux in a moving phase bin as histograms in \figref{fig:2ddiffhisto}.
We used \num{100} intensity bins, and we oversampled the phase bin
direction four times, resulting in \num{320} overlapping phase
bins each \num{0.0125} wide in phase.
There is larger spread in the \scshort data, and both the forward scattering peak as well as the egress are less visible than in the \lcshort data.
As expected from the analysis in \secref{subsec:sececlipse}, there are no signs of a secondary eclipse in either plot.
Besides features already investigated above, we see no features apart
from those explained by the 1-D model.

\section{Cloud model}
\label{sec:model}

\subsection{Model for the \lcshort data}
\label{subsec:model_lc}
The \keplersat \lclong (\lcshort) data are fitted by employing the one-dimensional model of \citet{brogi2012}.
Since the \lcshort data have an insufficient number of points per transit,
it is unrealistic to fit individual eclipses; we therefore only fit phase-folded and binned \lcshort data with this model, realising that we may average events that differ by more than just the transit depth.

Because of the improved data reduction and the much larger dataset
available, we also compared the \referee{best-fit values} and uncertainties derived here with those of our previous work.

\subsection{Model for the \scshort data}
\label{subsec:model_sc}
The analysis of the morphology of the \sclong \referee{data} (\scshort) (see \secref{subsec:primeclipse} and \figref{fig:sapbinned}) reveals residual structures after subtracting a properly scaled 1-D model.
This suggests that this model is not sufficient to describe the full morphology of the \tgtstarshort light curve, when observed with a higher time resolution.

\begin{figure}
  \centering
  \includegraphics[width=\linewidth]{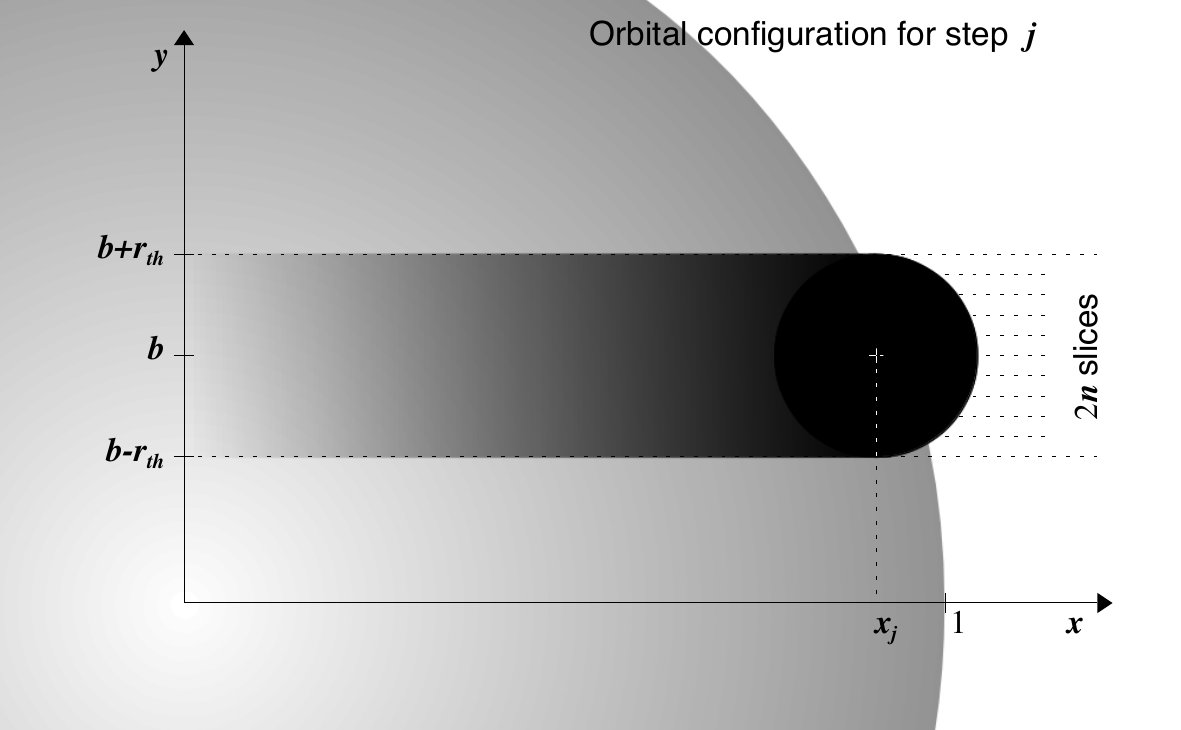}
  \caption{Geometry \referee{of} the two-component, 2-D  model.
  $x$ is along the planet candidate orbit, $y$ is perpendicular to that.
  $r_\mathrm{th}$ is the radius of the opaque core.
  \label{fig:model_geometry}}
\end{figure}

In an attempt to better fit the \keplersat \scshort data, we developed an improved cloud model by accounting for the vertical extent of the cloud, and by splitting the cloud structure into an opaque component for the dust around the planet, and an optically thin, exponentially-decaying cloud of dust following it.
When not accounting for the vertical extent of the cloud, the impact parameter derived from fitting the data is an \emph{effective} impact parameter, meaning that it is averaged over the non-uniform brightness of the stellar disk occulted by the cloud of dust.
This could partially explain why the impact parameters for \emph{deep} and \emph{shallow} data differ (see \tabref{tab:orbitparms}).
In addition, it is plausible that -- at least for the strongest outbursts -- the amount of dust ejected from the planet is sufficient to create an opaque envelope of material, which could also explain why the maximum transit depth seems to be limited to \SIrange{1.2}{1.4}{\percent}.
Indeed, it has been pointed out that the cloud should consist
of at least two component with different properties for the dust \citep{budaj2013}.

As a trade-off for the higher number of free parameters, we neglected the scattering component in this new model.
This is justified by the much lower photometric precision (by a factor of $\sim 45$) of the \keplersat \scshort, unfolded and unbinned data, with respect to the phase-folded and binned \lcshort data.
The scattering properties are almost exclusively constrained by the small peak just before ingress, which is completely buried in the noise in these data, as shown in \figref{fig:2ddiffhisto}.
The egress part of the light curve is also affected by scattering, but it is degenerate with the properties of the tail, meaning that by changing the exponential decay of the tail we could mimic the effects of a forward-scattering component.
Therefore, although this model may miss some of the physics involved, our aim is not to derive physical parameters or to compare them to the previous study, but to better understand the basic geometry of the cloud, and therefore we only focus on its structure.

The 2-D model shares the mathematical background of the previous 1-D code, i.e.\ it generates the light curve by convolving the profiles of the dust cloud and the stellar disk. 
The vertical extent of the cloud is described by 2$n$ slices centred on the position of the planet candidate.
The model is parametrised via the radius of the opaque, circular cloud of dust ($r_\mathrm{th}$), the scale-length of the exponentially-decaying, optically thin tail of dust ($\lambda_\mathrm{tail}$), and the impact parameter of the transit ($b$).
All these quantities are expressed in units of the stellar radius ($R_\mathrm{S}$), and both vertical and horizontal directions are quantised with the same step size $\Delta r$.
This is defined by subdividing the total length of the orbit into $m$ steps.
For a semi-major axis of $a = 4.31$, we have $\Delta r = 2\pi a/m \approx 27.1/m$.
We denote the vertical direction with $y$ (i.e.\ perpendicular to the orbit) and the horizontal direction with $x$ (i.e.\ along the orbit).
The curved path of the planet across the stellar disk for $b \ne 0$ is approximated with a straight line.

\Figref{fig:model_geometry} shows the geometry and the main quantities
involved in this new 2-D model.
The position along the orbit is traced through the vector $\vec{x} = [x_1, x_2, \cdots , x_m]$, in the interval $(-\pi a,\pi a)$, which is related to the orbital phase $\varphi$ via $\vec{\varphi} = \vec{x}/a$.
The zero point for the orbital phase coincides with the centre of the star and -- for $j=1$ in \eqnref{eq:totallc} -- with the position of \tgtshort.
At any time, the centre of the planet candidate, which also coincides with the centre of the opaque part of the cloud, is placed at $x=x_j$ and $y=b$.
The vertical position of the slices is defined by the vector $\vec{y}_\mathrm{cloud} = [y_1, y_2, \cdots, y_{2n}]$, which is centred on $y=b$ and in the interval $(b-r_\mathrm{th},b+r_\mathrm{th})$.
For each slice $i$, the absorption properties of the cloud are defined by
\begin{equation}
  \mathcal{C}_i(\vec{x}) = \left\{ 
    \begin{array}{ll}
      1 & \mbox{ for } |\,x\,| \le x_{\mathrm{th},i} \\
      \mathrm{e}^{-\vec{x}/\lambda_\mathrm{tail}} & \mbox{ elsewhere }
    \end{array}
    \right. ,
\end{equation}
where $\vec{x}_\mathrm{th} = \sqrt{r^2_\mathrm{th} - (\vec{y}_\mathrm{cloud}-b)^2}$ denotes the intersections of the orbital vector $\vec{x}$ with the opaque disk of dust, for each of the $2n$ slices.

The stellar profile is computed by assuming a quadratic limb-darkening law, meaning that the intensity of the stellar disk is expressed as
\begin{equation}
  \label{eq:int_mu}
  I_\mathrm{S}(\mu) = 1 - u_1(1 - \mu) - u_2(1 - \mu)^2,
\end{equation}
where $u_1, u_2$ are the quadratic limb-darkening coefficients for a star of similar properties as \tgtstar \citep{claret2011b}, while $\mu$ is the cosine of the angle between the line of sight of the observer and the normal to the stellar surface.
Therefore, $\mu$ is a function of two variables (the $x, y$ coordinates), and it is meaningful only for points inside the stellar disk.
For a given slice $i$, the intersection between the orbital vector $\vec{x}$ and the stellar disk (i.e.\ the edge of the star) is given by 
\begin{equation}
  x_{\mathrm{star},i} = \sqrt{1 - y_{\mathrm{cloud},i}^2} \, .
\end{equation}
The stellar brightness profile is therefore
\begin{equation}
  \mathcal{S}_i(\vec{x}) = \left\{ 
  \begin{array}{ll}
  I_\mathrm{S}(\vec{x},y_{\mathrm{cloud},i}) & \mbox{ for }  |\,x\,| \le x_{\mathrm{star},i}\\
  0 & \mbox{ elsewhere}  
  \end{array}
  \right. .
\end{equation}
The previous relation can be expressed explicitly by substituting 
\begin{equation}
  \mu = \sqrt{1 - \vec{x}^2 - y_{\mathrm{cloud},i}^2}
\end{equation}
in \eqnref{eq:int_mu}.
The total light curve is finally given by convolving the stellar and the cloud profiles for each slice, which is
\begin{equation}
  \label{eq:totallc}
  F(\vec{x}) = 1 - \sum_{i=1}^{2n} \sum_{j=1}^{m}\frac{\mathcal{S}_i(\vec{x})\, \mathcal{C}_i(\vec{x}-x_j)}{\mathcal{S}_\mathrm{tot}}.
\end{equation}
The normalisation \referee{factor} $\mathcal{S}_\mathrm{tot}$ (the total flux from the star) is precomputed by discretising the full limb-darkened stellar disk with the same step size as for the model, and summing over all pixels of the matrix.
It is therefore a much more straightforward normalisation than in our previous 1-D model, where the stellar profile for $b=0$ had to be normalised such that the sum of its elements was equal to unity.

In this two-dimensional model, fractional pixel coverage is also taken into account via a linear approximation.
This is particularly important for very small $r_\mathrm{th}$, or when
$b$ approaches ($1 + r_\mathrm{th}$), and prevents us from using a
too-small (and computationally demanding) $\Delta r$.
We validated our model with the \citet{mandel2002} transit code by choosing a very small value for $\lambda_\mathrm{tail}$, which is equivalent to neglecting the optically thin part of the cloud.
For an optimal value of $m = \num{3000}$, found via trial and error, the two models differ by less than \num{e-5}, which is at least \referee{two} orders of magnitude better than the accuracy of the \keplersat \sclong data. 

\subsection{Fitting of LC data}
\label{subsec:fit_lc}
The updated parameters of \tgtshort, as derived from the fitting of \num{15} quarters of \keplersat \lcshort data, are listed in \tabref{tab:orbitparms}.
For deep and shallow transits, we observe a better mixing of the Markov-chain Monte Carlo (MCMC) chains (quantified following \citet{gelman1992}) than in the previous work, which is expected from the much larger amount of data \referee{used}.
This also results in more symmetric posterior distributions and less correlation between \referee{the} parameters. 

\subsection{Fitting of SC data} 
\label{subsec:fit_sc}

\begin{figure}[!th]
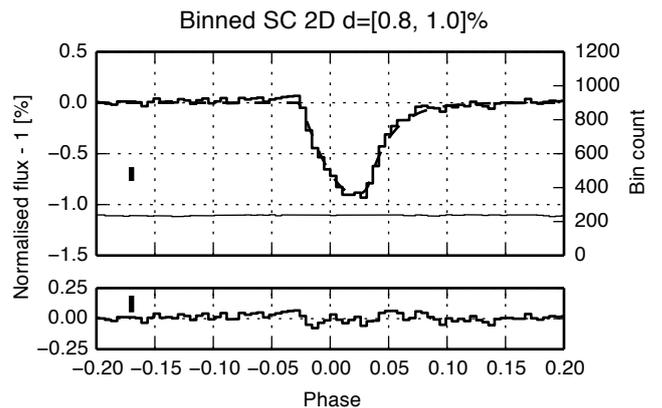
 
  \centering
  \includegraphics{{{\scorbtwodplotpath sap_sc_2d_binned_normd_nbin192_d0.8_1.0_with_residual2_model}.pdf}}
  \caption{\referee{Phase-folded, normalised flux binned in 192 bins for \sclong data for transits with depths from \SIrange{0.8}{1.0}{\percent}, as in \figref{fig:sapbinned}, but instead using the 2-D model to fit the data.
  Even though the 2-D model does not include forward scattering, we observe an improved fit and reduced residuals, with a \SI{16}{\percent} lower residual RMS of \num{3.1e-4}.}
  \label{fig:sapbinned2dmodel}}
\end{figure}

\referee{To compare our 2-D model with the 1-D model, we fit it to the subset of deep transits (see \tabref{tab:orbitparms}), and perform a similar analysis as in \secref{subsec:primeclipse}.
Even though we excluded the forward scattering from the 2-D model, the residuals are reduced as compared to the 1-D model, as shown in \figref{fig:sapbinned2dmodel}, indicating a better match to the data.
}

\referee{Subsequently}, we selected \num{213} light curves sampled in \scshort mode with
transit depths greater than \SI{0.5}{\percent}, no discontinuities in
the data and sufficient photometric precision to fit the 2-D model to the individual transits \referee{to be used for a per-orbit analysis}.
Each individual orbit contains \num{447} points. 

We fit \referee{the geometry of the dust cloud individually for} each light curve  by first performing a least-squares fit with a grid of widely spaced parameter values.
The set of parameters corresponding to the minimum $\chi^2$ is then used as input for a single MCMC chain of \num{50000} steps.
This assures that the chain is already started in the proximity of the global $\chi^2$ minimum, and does not get stuck in a local minimum. 

By performing the analysis with all free parameters, we notice that a large fraction of the MCMC chains fail to converge.
However, by investigating those chains that do converge, we observe that all transits are consistent with the same impact parameter ($b = \num{0.6\pm0.1}$).
This suggests that our two-component, two-dimensional model is a better approximation for the cloud of dust, and does measure a \emph{true} impact parameter, as opposed to the previous one-dimensional model.
Therefore, we fixed the impact parameter to \num{0.6} and recomputed the MCMC chains.

To verify that the model parameters are not degenerate, we computed artificial light curves by:
\begin{enumerate}
\item Assuming a small tail ($\lambda$ = 0.3) and only varying the opaque part;
\item Fixing $r_\mathrm{th} = 0.01$ and only varying the exponential tail.
\end{enumerate}
We added noise based on the photometric precision of the \keplersat \scshort data, and fit these simulated data with the method described above.
In both cases the retrieved parameters are compatible with the
single-parameter distributions given as input, and we do not obtain a
mixing between the two model components, as is observed in the real data.
This suggests that the two components are indeed present in the data.

\begin{figure}
  \centering
  \begin{subfigure}[b]{\linewidth}
    \includegraphics[trim = 0 10 0 0, clip=true]{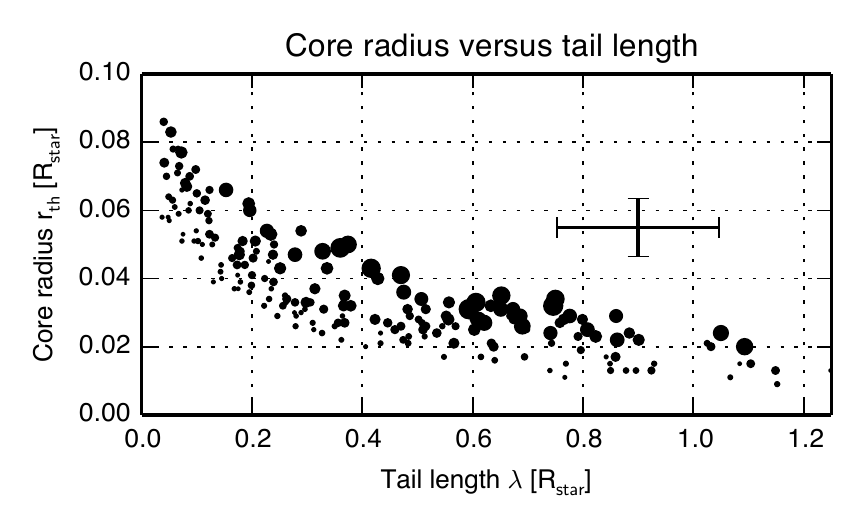}
  \end{subfigure}
  \begin{subfigure}[b]{\linewidth}
    \includegraphics[trim = 0 10 0 0, clip=true]{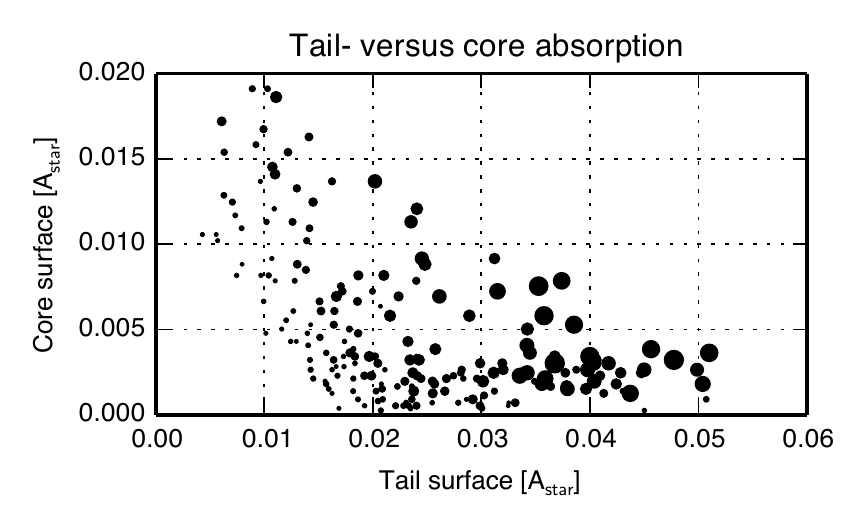}
  \end{subfigure}
  \begin{subfigure}[b]{\linewidth}
    \includegraphics[trim = 0 10 0 0, clip=true]{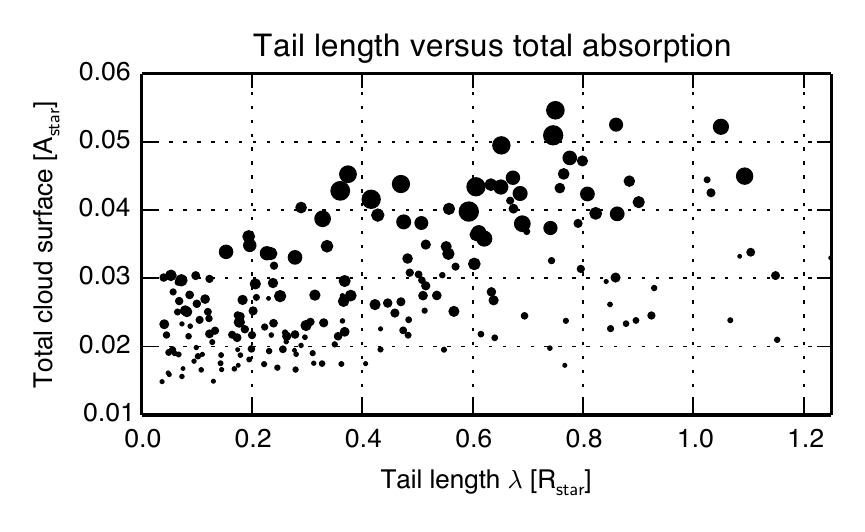}
  \end{subfigure}
  \caption{Deep ($>$\SI{0.5}{\percent}) \sclong transit \keplersat data fitted with the two-component 2-D model.
  The point \referee{surface} scales with the transit depth \referee{squared (for clarity)} as determined in \secref{subsec:primeclipse}.
  \referee{The top plot shows the best-fitting model parameters $r_\mathrm{th}$ and $\lambda$ for each individual \scshort transit.
  The correlation in this plot is largely due to the model (see text).}
  We show the effective area of the tail versus the size of the opaque core (\emph{middle}), and the tail length versus total absorption (\emph{bottom}).
  We observe no clear relation between the tail or the core absorption, while the transit depth scales with the total absorption.
  Very deep transits seem to require a strong tail component, \referee{as is evident by the cluster of points in the bottom-right corner of the middle panel.}
  The void in the triangular lower-left corner in the middle panel is due to the exclusion of shallow transits, \referee{where iso-transit depth lines run diagonally.}
  \label{fig:sc_tail_corr}}
\end{figure}

In our model the size of the opaque part also drives the vertical extent of the cloud.
This means that, if the tail covers a certain effective area (integrated over the $x$ and $y$ direction), its specific length $\lambda$ has to change as a function of the size of the opaque core.
Larger $r_\mathrm{th}$ means a wider cloud, which requires a faster decay in order to maintain approximately the same effective area.
\referee{Simply plotting $r_\mathrm{th}$ versus $\lambda$ therefore shows a correlation due to the model, and tells little about the system itself (see \figref{fig:sc_tail_corr}).}
For this reason we plot the absorption by the core as $\pi \, r_\mathrm{th}^2$ and of the tail as $2 \, r_\mathrm{th} \, \lambda_\mathrm{tail}$.

The results of the \sclong transit analysis are shown in \figref{fig:sc_tail_corr}.
\referee{Besides plotting the model parameters}, we also show the absorption of the core versus the absorption of the tail \referee{and} the tail length versus the total absorption \referee{since these better describe the physics of the system}.
The total absorption is highly correlated with transit depth (\referee{bottom panel;} as determined in \secref{subsec:primeclipse}) which is expected: more material will yield a deeper minimum.
We find no relation between the tail length and the transit depth or total absorption \referee{(bottom panel)}.
Although shallow transits are evenly distributed between the core and tail absorption, deeper eclipses appear to have most absorption in the tail, and not in a disk-shaped, opaque coma \referee{(bottom-right corner of middle panel)}.

Our model is certainly not the only possible geometric description of \tgtshort, and it is unclear whether the two components we propose are an accurate physical description of the system.
One could, for example, fix the vertical extent of the cloud to allow
for a variable optical depth for the core with an exponentially decreasing tail.
The main point is that the data suggest the necessity of at least two
independent components for the cloud, in agreement with \citet{budaj2013}.

\section{Conclusions}
\label{sec:conclusions}

We have manually de-correlated and de-trended \num{15} quarters of
\keplersat data, of which three quarters were observed in \sclong
mode for \tgtstar, and investigated statistical constraints on system
dynamics as well as a per-orbit analysis using three quarters of \scshort data.

We find two quiescent spells of \tgtshort of $\sim$\num{30} orbits each around orbit \num{50} and \num{1950} where the average transit depth is \SI{0.1}{\percent}.
These two periods appear to be followed by periods where the transit depth is \SIrange{0.1}{0.2}{\percent} deeper than the average transit \referee{depth}.
Additionally, we find times at which the transits show on-off-like behaviour, where $>$\SI{0.5}{\percent} deep eclipses are followed by hardly any eclipse at all.
Aside from these isolated events, we find no significant overall correlation between consecutive transit depths, nor between transit depth and \referee{consecutive} egress depth.
This implies that the majority of the dust does not survive a single
orbit, and that the process underlying the dust generation occurs
erratically.
The independence of transit depth and consecutive egress depth implies that an opaque dust cloud yielding a deep transit does not survive to form part of a more tenuous and stretched-out dust cloud during the next eclipse\referee{, in agreement with \citep{perezbecker2013}}.

Furthermore, we put an upper limit of \num{4.9e-5} on the secondary eclipse.
This constrains the radius of a planet candidate to less than \SI{4600}{\kilo\meter}, or one Earth radius for an albedo of $\sim$\num{0.5}.

We find a significant discrepancy when fitting our previous 1-D model to the \scshort data around egress.
Our improved model adds a second dimension and represents the
dust tail with two components, which better fits individual \scshort
transits than the old 1-D model.
We verified that the two components are not degenerate in the model and are data-driven.

Our results suggest that a 1-D, exponentially decaying dust tail is insufficient to represent the data.
We find that deep transits have most absorption in the tail, and not in a disk-shaped, opaque coma, but the transit depth or total absorption show no correlation with the tail length.
Although our model is only one possible interpretation of the cloud structure, there is also qualitative evidence \citep{budaj2013} of the need of at least two components.
We anticipate that \referee{models including realistic physics and geometry} are required to fully understand this peculiar system. 

\begin{acknowledgements}
The authors would like to thank Ernst de Mooij for his constructive
discussion concerning the modelling of this object.

This paper includes data collected by the \keplersat mission. Funding for the \keplersat  mission is provided by the NASA Science Mission directorate.

All of the data presented in this paper were obtained from the \emph{Mikulski Archive for Space Telescopes} (MAST). STScI is operated by the Association of Universities for Research in Astronomy, Inc., under NASA contract NAS5-26555. Support for MAST for non-HST data is provided by the NASA Office of Space Science via grant NNX09AF08G and by other grants and contracts.
\end{acknowledgements}

\bibliographystyle{aa}
\bibliography{tvwerkhoven}

\end{document}